\documentstyle[12pt]{article}

\begin{document}

\baselineskip=18.6pt plus 0.2pt minus 0.1pt

\makeatletter
\@addtoreset{equation}{section}
\renewcommand{\theequation}{\thesection.\arabic{equation}}

\begin{titlepage}
\title{
\hfill\parbox{4cm}
{\normalsize HEPL/2000\\
hep-ph/0006248}\\
\vspace{1cm}
    On Geometric  Engineering  of Supersymmetric Gauge Theories 
}
\author{
Adil {\sc Belhaj}\thanks{{\tt ufrhep@fsr.ac.ma -Talk  given at  Workshop on Non commutative geometry, Superstrings  and Particle Physics,  16-17 June 2000, Rabat, Morocco  }}
{} 
\\[7pt]
{\it  High Energy Physics Laboratory,  Faculty of sciences ,  Rabat ,  Morocco}
}

\maketitle
\thispagestyle{empty}

\begin{abstract}
 We  present the basic ideas of   geometric engineering of  the supersymmetric quantum field theories  viewed  as  a low energy limit of  type  II strings and F-theory on singular  Calabi Yau  manifolds.  We first give the  main lines of toric geometry as it is a powerful technique to deal with compact complex manifolds. Then we introduce  mirror symmetry  which plays a crucial role in the study of superstring dualities and  finally we give elements on Calabi Yau singularities.  After that we  study the geometric engineering of $N=2$ supersymmetric gauge theories in six and four dimensions. Finally we make comments regarding $ N=1$ SYM in four dimensions.
\end{abstract}
\end{titlepage}
\newpage
\section{Introduction}
 Over the  few past years we have learned how many  non trivial    supersymmetric  QFT  are obtained from singular limits of  type II strings, M-theory and F-theory compactifications by using  the geometric engineering method introduced \cite{KKV} and developped  by Kazt, Mayr and Vafa  in \cite{KMV},  see also \cite{{BFS},{BFaS},{BS}}. The basic tools of  this tricky  method are toric geometry of ADE singularities of  the K3 surface and local mirror symmetry. In  this method, the  complex and Kahler deformation parameters of the singularities are related to physical parameters in the low energy  limit of string compactifications. The most familiar example is  type IIA string on K3 near ADE singularities, which believed   to be dual to heterotic string on $T^4$. In this case the low  energy limit is described   by supersymmetric  theories in six dimensions with ADE gauge groups. The moduli space of gauge invariant vacua of these models is just the moduli space of K3,  which may be  viewed as the moduli space of wrapped D2 branes on two cycles of K3.\\
 In this communication   we focus our attention on the study of type II  superstrings  compactification on  ADE hypersurface singularities. In particular we study the embedding  of $4D$ supersymmetric  QFT in type IIA string compactification on Calabi Yau threefolds near the ADE singularities. We  also  discuss  the geometric engineering of the interesting  case of  $N=1$  supersymmetric gauge  theories, obtained from F-theory on singular elliptic Calabi Yau manifolds \cite{BS}. 
\section{Toric geometry and Calabi Yau singularities} 
Here we review briefly some basic facts about toric geometry useful for  the study of superstring compactifications \cite{LV}. Roughly speaking,  toric geometry  concerns $n$  complex dimensional  manifolds  which can be represented  by a polytope $\Delta$  of the $n$ dimensional $Z^n$ hypercubic lattice of $R^n$.  Instead of using  direct  complex analysis methods for complex manifolds, it is interesting to apply technics of toric geometry.  Toric geometry is a  valuabe tool for the discussion of  geometric  properties of Calabi Yau manifolds which are important in the context of string compatifications and  Calabi Yau fiber bundles  involved in F-theory - heterotic strings  duality. \\
Simple example of toric varieties are given by  weighted projective spaces $(WCP^n)$. These spaces can be defined as
\begin {equation}
WCP^n = {C^{n+1}-\vec 0\over {C^*}},
\end{equation}
with the ${C^*}$ action
\begin {equation}
 {C^*}: x_i   \to \lambda ^{q_i} x_i , \quad i=1,2,\ldots ,n+1.
\end{equation}
 The leading $ CP^1$ example parametrized  by $\{(x_1,x_2)/(x_1,x_2)=(\lambda x_1,\lambda x_2),\lambda \in {C^*}\}$ is just the complex line which is  know to be isomorphism to the real two sphere $ S^2 \approx SU(2)/U(1)$. A less trivial example is given by   
 $ WCP^2(2,3,1)$. In this case, the  equivalence relation (2.2) becomes 
\begin {equation}
 (x_1,x_2,x_3)\to (\lambda^2 x_1, \lambda^3 x_2, \lambda x_3).
\end{equation}
 This relation can be encoded in  triangle  in $ R^2$, with the following three vertices  $v_{x_1}=(-1,0)$, $v_{x_2}=(0,-1)$ and $v_{x_3}=(2,3)$ in  $ Z^2$, such that 
   \begin {equation}
 2 v_{x_1}+3v_{x_2}+ v_{x_3}=0,
\end{equation}  
where the coefficients of $ v_{x_i}$ are the powers of $\lambda$ in eq (2.3). More general  d-dimensional toric manifolds  are  generalizations of these weighted projective spaces  which are  defined as
              
\begin {equation}
V_{\Delta}^d = {C^{k}-U\over {C^*}^r},
\end{equation}
where  the $U$ set and the  $ C^*$ action  are given by:
\begin {equation}
 {C^*}^r: x_i   \to \lambda ^{q_i^a} x_i , i=1,2,\ldots ,d+r ; a=1, 2,\ldots ,r
\end{equation}

\begin {equation}
 U= \cup _I \{ (x_1,\ldots x_{d+r}), \quad x_i=0 \quad   for  \quad   i  \in I \}.   
\end{equation}
 The toric manifold (2.5) extend the complex projective spaces $ WCP^n$ in the sense that instead of removing the origin,   one removes the set $ U$  and  takes the quotient by   the  $ C^*$ actions. D-dimensional toric manifolds $V_{\Delta}^d $ have many   remarkable  properties  one of them  is that they may  be encoded in toric diagram $ \Delta$  of $ k=d+r$  vertices $ v_i$ embedded in the  $Z^d$ lattice  such that 
    \begin {equation}
  \sum \limits _{i=1}^{d+r} q_i^a v_ i=0, \quad  a=1,\ldots,r.
\end{equation}
In  these eqs (2.8), the $q_i^a$'s are  the Mori vectors defining the intersection matrix of  divisors of toric manifolds $V_{\Delta}^d $. Note in passing  that toric  geometry is intimately related to  $2d$ $N=2$ supersymmetric   sigma models. In the case 
 where the  target space  is  $V_{\Delta}^d $ ,   then  the  charges $ q_i^a $ are  interpreted  as the charges of the matter fields $(x_i)$, and eqs (2.8) is linked to   the D- flatness eqs  of $2d$ $N=2$   gauge theory  describing the flat direction as shown here below   
  
     \begin {equation}
  \sum \limits _{i=1}^{d+r} q_i^a |x_i|^2=R_a.
\end{equation}
 In this eqs   $ R_a$'s  are  the  Fayet Iliopoulos (FI) coupling parameters which describe Kahler parameters of toric manifolds.  Note also that  the first chern class of these spaces is proportional to $\sum \limits _{i,a} q_i^a$ \cite{HV}.  For $\sum \limits _{i,a} q_i^a=0$, the $V_{\Delta}^d  $ become  Calabi Yau manifolds. This means physically that the $ N=2$ supersymmetric theories flow to $ N=2$ SCFT \cite{{HV},{HIV}}.\\
Another important tool of toric geometry is mirror symmetry. The latter is a symmetry  which transforms into each other  Kahler and complex structres of  complex d- dimensional Calabi Yau  manifolds $ M$ and $W$.   A mirror  pair  has Hodge numbers  satisfying the mirror relations.    
   \begin {equation}
 \begin{array}{lcr} 
h^{1,1}(M)= h^{d-1,1}(W)\\
h^{d-1,1}(M)=h^{1,1}(W),
 \end{array}
\end{equation}
 this means that the complex moduli space of $M$ is identical  to the Kahler moduli space of $W$ and vice  versa. Mirror symmetry  plays a central role in the study of  type II superstring compatifications and in the determination of the moduli space of vacua.  This transformation  can be  viewed as a generalization of  T-duality in type II strings compactification on Calabi Yau manifold \cite{LV}.  It would be interesting to note that the mirror symmetry has played a crucial role in the developement in superstrings and QFT dualities and in the obtention of exact results. For example, l'absence  of type IIA dilaton in the vector multiplet, has been exploited to derive exact solution in the Coulomb branch of $ N=2$ QFT in four dimensions by using mirror symmetry. 
\subsection{ADE hypersurfaces} 
Toric varieties  may  have  singularites, which are  very important  in the understanding the  non perturbative  solutions of gauge theories. Some of these singularities,  which have toric realizations,  are given by the so called  ADE singularites: 
\begin {equation}
\begin{array}{lcr}
A_n: xy+z^n=0\\
D_n: x^2+y^2z+z^{n-1}=0\\
E_6: x^2+y^3+z^4=0\\
E_7: x^2+y^3+yz^3=0\\
E_8: x^2+y^3+z^5=0.\\
\end{array}
\end{equation}
These equations describe complex surfaces embedded in $C^3$ with coordinates $x,y,z$. Each of them has a singularity at $ x=y=z=0$. ADE singularities  may be resolved  either by deforming the complex structure  or  the  Kahler one to obtain a smooth manifold. Kahler deformation consists to blow up  the singular point by intersecting 2-spheres ranged according to the  Dynkin diagram  of   ADE Lie algebras. The intersection matrices of the blowing up 2-spheres $C_2$ , of the resolution  of the ADE singularities, are given by the Mori vectors  $q^a_i$ (eq ( 2.4-5)) ,  which up to sign, coincides  with the Cartan matrices $K_{ij}$ of the $ADE$ Lie algebras. This nice connection between singularities and Lie algebras  plays an important role in the geometric engineering of the $ N=2$ supersymmetric quantum field theory in four dimensions  obtained  from  type II strings compatification on local Calabi Yau threefolds \cite{KMV}, and in the geometric engineering of $ N=1$ models from F-theory compactification on elliptic Calabi Yau manifolds \cite{BS}. 
\section{ Geometric  engineering   of $ N=2$   QFT in four dimensions }
Geomeetric engineering of $ 4D$ $ N=2$ supersymmetric quantum field theories is  a geometrical method allowing to get the relevent moduli from type IIA string compactification on  local  Calabi Yau threefold $M_3$  with  ADE singularites. In this  method  $M_3$ is  realized as a local K3 ( ALE space) fibered over a  base which may be thought of as a 2-sphere or a collection of intersecting 2-spheres. The gauge  fields  of the QFT is obtained from D2 branes  wrapped on the 2 cycles of  the singularities of the fiber $K3$ and 
  the matters are  given by non trivial geometry on the base of $ M_3$. The physical parameters of the field theory are related to the moduli space of   both the fiber (F) and the base (B) of $ M_3$. The gauge coupling  $g$ is proportional to the inverse of the square root of the volume of the base $ V(B)$, i.e
$$ V(B)=g^{-2}.$$ 
  Before   giving   the main steps in getting $ 4D$$ N=2$ from IIA string on $ M_3$. Let us begin   by describing  type IIA  compactification on local K3 with ADE singularities.    
\subsection{$N=2$ in six dimensions}
  Type IIA on K3 near  ADE singularities  give a $6D$ $ N=2$  supersymmetric  gauge theory with ADE gauge symmetries. To fix the  ideas  suppose  for simplicity that $K3$ has a $su(2)$ singularity. The local geometry of this  background is  described by the complex equation:
$$ xy=z^2$$ 
 Type IIA on K3  with $su(2)$ singularity   gives a  $ N=2$    $SU(2)$ gauge theory in six dimensions. In this case  D2-branes wrapping around the blow  down  2-sphere    give two $W_\mu^{\pm}$   massless vector particles depending of the two possible orientations for the  wrapping. The  $W_\mu^{\pm}$   gauge fields are charged under the $U(1)$  gauge boson $Z_0^\mu$ obtained by decomposing the type IIA superstring 3-form in terms of the harmonic forms of   the vanishing  2-sphere .  Then  near an $A_1$ singularity of K3, we get   three massless vector particles $W_\mu^{\pm}$   and $Z_0^\mu$  which  altogether form an  $SU(2 )$ adjoint. We thus obtain a $N=2$ $SU(2)$ gauge theory  in 6 dimensions.  More genererally, if the single  vanishing 2-sphere is replaced by a collection of intersectiong  2-sphere according to the ADE Dynkin Diagrams, one get a $6D$ $ N=2$ sypersymmetric gauge theory  with ADE gauge groups.  
\subsection{$N=2 $ in four dimensions}
To  obtain QFT's in  four dimensions,  one has to  consider a  further compactification   on a one complex dimensional  base of $M_3$. If the (B)  is a  single 2-sphere, then  one gets  a $N=2$ pure $SU(2)$ Yang -Mills in 4 dimensions. To incorporate matter, we consider non trivial geometry on the base of $M_3$.  If we have a 2 dimensional locus with $SU( n)$  singularity and another locus with $SU( m)$ singularity and they meet to a point, the mixed wrapped 2-cycles will now lead to $(n ,m) $ $N=2$ bi-fundamental matter of the $SU(n)\times SU(m)$ gauge symmetry in four dimensions.  Geometrically, this means that the  base geometry  of $M_3$  is given by two  intersecting $ P^1$ 2-spheres  whose volumes $V_1$ and $V_2$  define the gauge coupling constant $g_1$ and $g_2$ of the $SU(n)$ and $SU(m)$ symmetries respectively. Note that we can  also engineer the adjoint matter. Moreover if we choose the base (B) as  a collection of intersecting 2-spheres  according to affine Dynkin diagrams, then one engineers  $ N=2$  superconformal  field theories  in  four dimensions.   \\ 
Geometric engineering of $ 4D $ $N=2$ QFT is really a tricky method to study $4D$ $ N=2$ QFT embedded in type IIA superstring theory. In this method, $4D$$ N=2 $ QFT's are represented by quiver diagrams where for each $SU$ gauge group factor we consider a node, and for each pair of groups with bi-fundamental matter, we connect the corresponding nodes with a line. These diagrams have a similar representation as the  ADE Dynkin diagrams of ordinary and affine simply laced Lie algebras. The developments obtained over the few last years are nicely described in this approach. In this regards, it is worthwhile to mention the  three following  due to this construction : \\
(i) The derivation  of exact solutions of  Coulomb branch  of $4D$ $N=2$ QFT which are obtained by help of local  mirror symmetry. \\
 (ii)   The classification   of  $4D$ $N=2$ superconformal theories  in terms of  affine ADE diagrams. This analysis is also valid in  the non simply laced   cases \cite{BFaS} \\
(iii) The gauge coupling space of these  superconformal field  theories is linked to  the moduli of flat connections on the torus. These  moduli  is interseting in the study  of the duality  bettween heterotic string  on elliptically fibered compact manifolds and F-theory, and in geometric  construction of $ N=1$ vacua .
\section{  Conclusion and  Discussion }
   We  conclude this communication  by discussing the geometric construction of $ N=1$ Yang - Mills in four dimensions. These models may be obtained in  terms  of F-theory on elliptic Calabi You fourfold. The latter is realized as elliptically  fibered K3, with affine ADE singularities,  fiber over a complex base space. If we choose  the base  a  $ P^2$, or two complex dimensional toric spaces $ F^n$ , this  gives  $ N=1$ Yang Mills in four dimensions with ADE gauge symmetries. Morevover we can also engineer $ N=1$ models  with non simply laced gauge symmetries  by using  the analysis of \cite{BS}. This analysis is based on toric realization of folding  method of ADE Lie algebras.  In that  our work we have  distinguich two possible toric realizations  depending on the action of the folding on the elliptic curve of the fiber K3. \\
 In this work we have studied  the  geometric  engineering of supersymmetric gauge theories obtained as a low energy limit of  type II strings on Calabi Yau manifolds  with  ADE singularities in the fiber K3.  It turns out  that  ADE singularities of K3 lead to appearance of corresponding gauge group in physics. Moreover these singularities have toric realizations, which are  related to Kahler Calabi Yau construction from $ 2D$ $ N=2$ sigma models. Thus it is natural to think about another type of   singularities and their gauge theories corresponding,  such as hyperkahler singularities  which are linked to $ 2d$ $N=4$ sigma models approach \cite{{BH},{BSa}}. These singularities  maight be used to derive new physics, not described  by a conventional gauge theory.\\
 {\bf Acknowledgement}\\
Adil Belhaj would like thank  to the organizers of the workshop on Non Commutative Geometry, Superstrings  and Particle Physics( 16-17 June  2000) Rabat, Morocco, and  the National Network for Theoretical Physics (NNTP) for hospitality. He would  also like  to thank E.H. Saidi for discussions.\\
This  work  supported by the programmm PARS  PHYS 27. 372/98 CNR.


\begin{thebibliography}{99}

\bibitem{KKV}
S.\ Katz  , A.\ Klemm and  C.\ Vafa, `` Geometric engineering of quantum field theories ,''
  {\em Nucl Phys.} {\bf B497} (1997) 173.
{{\tt hep-th/9609239}}.
\bibitem{KMV}
S.\ Katz, P.\ Mayr and  C.\ Vafa,
  `` Mirror Symmetry  and Exact Solution of $4D$ $N=2$ gauge theories ,''
  {\em Adv. Theor. Math. Phys.} {\bf 1} (1998) 53.
 {{\tt hep-th/9706110}}. 

\bibitem{BFS}
A.\ Belhaj, A. El. \ Fallah and E. H. \ Saidi ,`` On the affine $ D_4$  mirror geometry ,'' {\em CQG}{\bf 16} (1999) 3297 .


\bibitem{BFaS}
A.\ Belhaj, A. EL.\  Fallah  and E. H. \ Saidi , ``On the non-simply laced mirror geometries in type II strings ,''
  {\em CQG} {\bf 17} (2000) 515
\bibitem{BS}
A.\ Belhaj,  and E. H. \ Saidi in  progres

\bibitem{LV}
 N.C. \ Leung and C. \ Vafa,'' Branes and Toric Geometry,
  ''{\tt hep-th/9711013}.
\bibitem{HV}
 K.\  Hori,  and C.\  Vafa ''  Mirror Symmetry''
{\tt hep-th/0002222}.

\bibitem{HIV}
 K.\  Hori, A.\  Iqbal and C.\  Vafa,'' D-branes and Mirror Symmetry'',
{\tt hep-th/0005247}.
\bibitem{BH}
A.\ Belhaj,  and E. \ H.  Saidi,''  HyperKahler singularities in superstrings compactifiaction and $ 2d$ $ N=4$ conformal field theory '',
{\tt hep-th/0002205}.


  \bibitem{BSa}
A.\ Belhaj,  and  E.  H. \ Saidi,''  On HyperKahler Singularities '' ,{\tt hep-th/0007143}





\end{thebibliography}
\end{document}